\documentclass[prb,aps,espf,showpacs,twocolumn]{revtex4-1}
\usepackage{amsmath}
\usepackage{graphicx}
\usepackage{color}

\begin{document}

\title{ Distortion and electric field control of band structure of silicene}

\author{Gul Rahman$^{1}$}\email{gulrahman@qau.edu.pk}
\affiliation{$^1$Department of Physics,
Quaid-i-Azam University, Islamabad 45320, Pakistan}
\begin{center}
\hspace{2cm}\textbf{Europhys Lett.( 2014) Accepted}
\end{center}
\begin{abstract}
Density functional theory with local density approximation for exchange and correlation functional is used to tune the electronic band structure of silicene monolayer. The cohesive energy of free standing monolayer is increasing (decreasing) with external electric field (distortion). Electrons in silicene behave like  Dirac fermions, when the bond angle between the Si atoms is larger than $\sim 102^{0}$. Large distortions destroy the electronic structure of silicene and silicene is no longer a semi-metallic material, and the distorted silicene acts like an $n$-doped system. Electric field  opens a band gap around $K$ point in the Brillouin zone, which increases with electric field. The bond angle between the Si atoms is a key player to determine the presence or absence of Dirac cones in silicene. \end{abstract}


\maketitle

\section{Introduction}
\label{sec:intro}
In recent years, two-dimensional (2D) materials have gained intense interest not only from theory communities, but also from  experimental scientists. 2D materials are interesting due to their unique physical properties. For example 2D MoS$_{2}$ is a direct band gap semiconductor, whereas 3D MoS$_{2}$ is an indirect band semiconductor, while NbSe$_{2}$ retains its metallic character both in 2D and 3D structures.\cite{Erikson} In contrast 2D Ge is a poor metal, whereas 3D Ge is a semiconductor.\cite{Erikson}
Reduced dimensionality can sometimes help to reduce the defect formation energy of magnetic systems.\cite{Rahman}  Light elements (C, Si, Ge, e.g.)based 2D structures have very unique properties. Graphene, which is a 2D crystal with carbon atoms arrayed in the flat hexagonal lattice of graphite, has Dirac-like electronic structure with linear dispersion around the Fermi level ($E_{\rm{F}}$).\cite{Neto} Therefore, graphene is considered to be a host for Dirac type electrons, whose unusual properties have been studied extensively in graphene monolayers produced by mechanical   
exfoliation from graphite.\cite{Novoselov, Geim}

Silicene, which is a 2D buckled monolayer honeycomb structure of Si atoms, has also gotten much attention around the world.\cite{Novoselov, Geim}
It is demonstrated by first-principles calculations that low-buckled silicene is dynamically stable and has a linear electronic dispersion relation near $K$ points at the corner of the Brillouin zone (BZ).\cite{Takeda,Drum, Durgn,Zhang,Cahangirov} The stability of silicene can be understood from the nature of $sp$ hybridization. Although silicene is isoelectronic to graphene
,but Si has a larger ionic radius,
which promotes $sp^{3}$ type hybridization.  On the other hand,  $sp^{2}$ type hybridization
is energetically more favourable in C, whereas  in a 2D
layer of Si atoms, the bonding is formed by mixed $sp^{2}$ and
$sp^{3}$ hybridization. Hence, silicene is slightly buckled and 
such buckling creates new possibilities for manipulating the
dispersion of electrons in silicene and opening an electrically
controlled sublattice-asymmetry band gap,~\cite{ZNi} which is not possible in graphene due to sublattice-symmetry. The electronic $\pi$- and $\pi^{*}$-bands derived from the Si $3p_{z}$ orbital disperse linearly to cross at $E_{\rm{F}}$, and the silicene electrons behave as massless Dirac fermions. This unusual property of silicene puts it as a promising candidate for quantum spin Hall effects.\cite{Hall,Drum}  

Recent experiments have demonstrated  silicene on  different substrates~\cite{Ag,ZrB2,Ir} with a hope to utilize massless fermions in silicene. However, the results are somehow conflicting. Fleurence {\it {et al.}}\cite{ZrB2} have shown that buckled silicene on ZrB$_{2}$ has an energy gap at $E_{\rm{F}}$. The angle-resolved photoelectron spectroscopy (ARPES) experiment of $4\times 4$ silicene showed a linear band structure, which is considered to be the finger-print of Dirac fermions, and the Dirac point was measured to be 0.3 eV below $E_{\rm{F}}$.\cite{Ag1} In another experiment carried out by Chen and his co-workers,~\cite{Chen} the Dirac point was $\sim 0.5$ eV below $E_{\rm{F}}$. In both of these experimental work the Fermi velocity was close to the theoretical value of $\sim 10^{6}$ $m/s$.\cite{Cahangirov} On the other hand, density functional theory (DFT) calculations found linear dispersion only within a energy interval $\pm 0.40$ eV, which is smaller than the experimentally reported values.\cite{Ag1,Chen} In spite of these two pioneer experimental work, a very recent experimental study~\cite{Lin} on Landau level in silicene on Ag and absence of the characteristic signals was attributed to the Landau level, which does not agree with the previous experimental work of Vogt {\it{et al.}}\cite{Ag1} and Chen {\it{et al.}}\cite{Chen}

From the above comprehensive literature, it can be inferred that it is very essential to investigate the nature of Dirac point in silicene in different external perturbations, e.g., strain, pressure, electric field, etc.  Silicene can be deposited on different substrates for practical applications and its electronic properties can be perturbed by strain or interfacial effect. So, from applications point of view, we must investigate the desired properties against strain/distortion, which can be induced by substrate. This effect can not be ignored in real devices because there is always lattice mismatch whenever we grow a material on a substrate. In some cases, such distortion/strain is technologically beneficial and can be used to engineer the band structure of a material.~\cite{Rahman-CrP} Therefore, in this article, we inquire the electronic structure and cohesion of silicene under distortions and electric fields.

\section{Computational Method}
\label{sec:comp}
The present calculations are performed  in the framework of DFT ~\cite{DFT} using linear combinations of atomic orbitals (LCAO) as implemented in the SIESTA code. ~\cite{siesta} We used a double-$\zeta$ polarized (DZP) basis set for all atoms. We employed  the local density approximation (LDA) as parametrized by Ceperly and Adler, \cite{lda}  for the exchange-correlation functional.
We used standard norm-conserving pseudopotentials ~\cite{ps} in their fully nonlocal form .\cite{pss}  We found that 200 Ry was enough to converge the lattice parameters and band structure of bulk Si and monolayer of silicene. The
Brillouin zone integration was performed using  Monkhorst-Pack grids of $45\times 45\times 1$. 

The optimized lattice constant and buckling parameter $\delta$ of silicene are $3.83$\,\AA and $0.44$, respectively, which are comparable with the previous work.\cite{Drum} As mentioned in the introduction that when silicene is deposited on a substrate then it is expected that silicene may distort slightly, therefore, we considered different distortions. We mainly distort the bond angle between the Si atoms, that will also change the bond length of Si-Si. In such distortions, the lattice parameters of silicene were kept at their optimized values. The distorted systems are denoted by $S_{d}\;({d=}$0-5).  Here $S_{0}$ corresponds to pristine silicene. The bond angles(lengths) of all the studied systems are shown in Table~\ref{angle}.  
\begin{table}[!h]
\label{angle}
\centering
\caption{Bond angle(in degree) and length(in \AA) between Si atoms in silicene. The first column shows different systems $S_{d}$ ($d=0-5$).}
\begin{tabular}{cp{0.3cm}cp{0.3cm}cp{0.3cm}cp{0.3cm}c}
\hline
\hline
System && bond Angle && bond length &&  \\
\hline
S0&&      116.28&&                        2.250&&\\
S1&&      112.36&&                        2.300&&\\
S2&&      107.85&&                        2.367&&\\
S3&&      102.93&&                        2.446&&\\
S4&&      98.00&&                         2.530&&\\
S5&&      92.77&&                         2.640&&\\

\hline 
\hline 
\end{tabular}
\label{angle}
\end{table}

\section{Results and Discussion}
\label{sec:result}

It is essential to discuss the cohesive energy of silicene under different strain and electric field. First, we calculated the cohesive energy of bulk Si. The calculated $E_{c}$ of bulk Si is $\sim 4.51$ eV, which is comparable with the experimental values $4.62 - 4.88$ eV/per atom. Our calculated $E_{c}$ does not include a correction for the zero point energy which is about 0.103 eV~\cite{Drum}. Once we confirmed that our bulk $E{_c}$ agrees with previous work,  we then calculated $E_{c}$ of silicene which is $\sim 4.85$ eV. This value is also comparable with calculation using other functional and computational codes.\cite{Drum} This shows that bulk Si is 0.34 eV per atom more stable than silicene. If we include the previously~\cite{Drum} LDA calculated  zero point correction then silicene is about 0.24 eV per atom higher than bulk Si.
\begin{figure}[]
\includegraphics[width=0.23\textwidth]{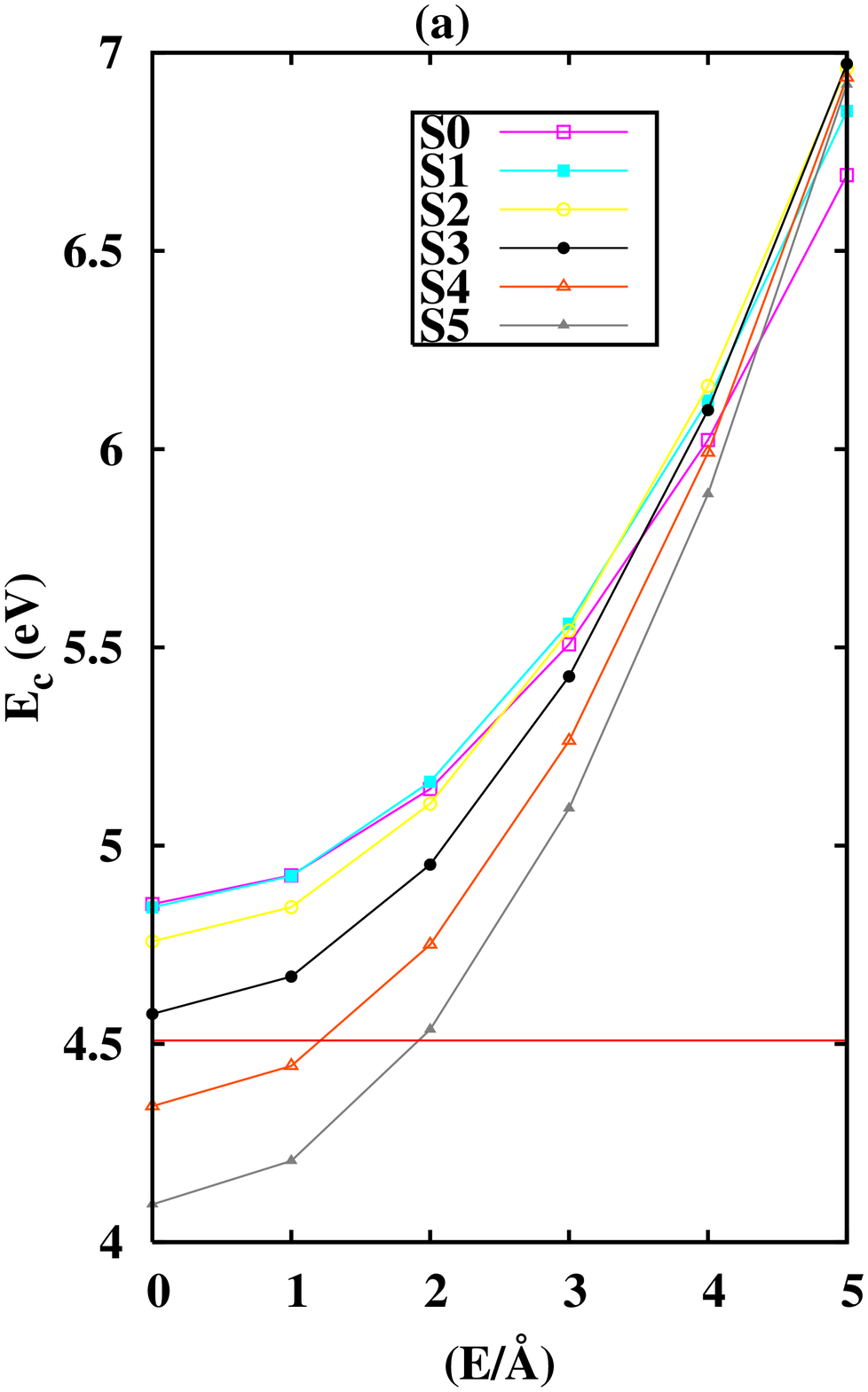}
\includegraphics[width=0.23\textwidth]{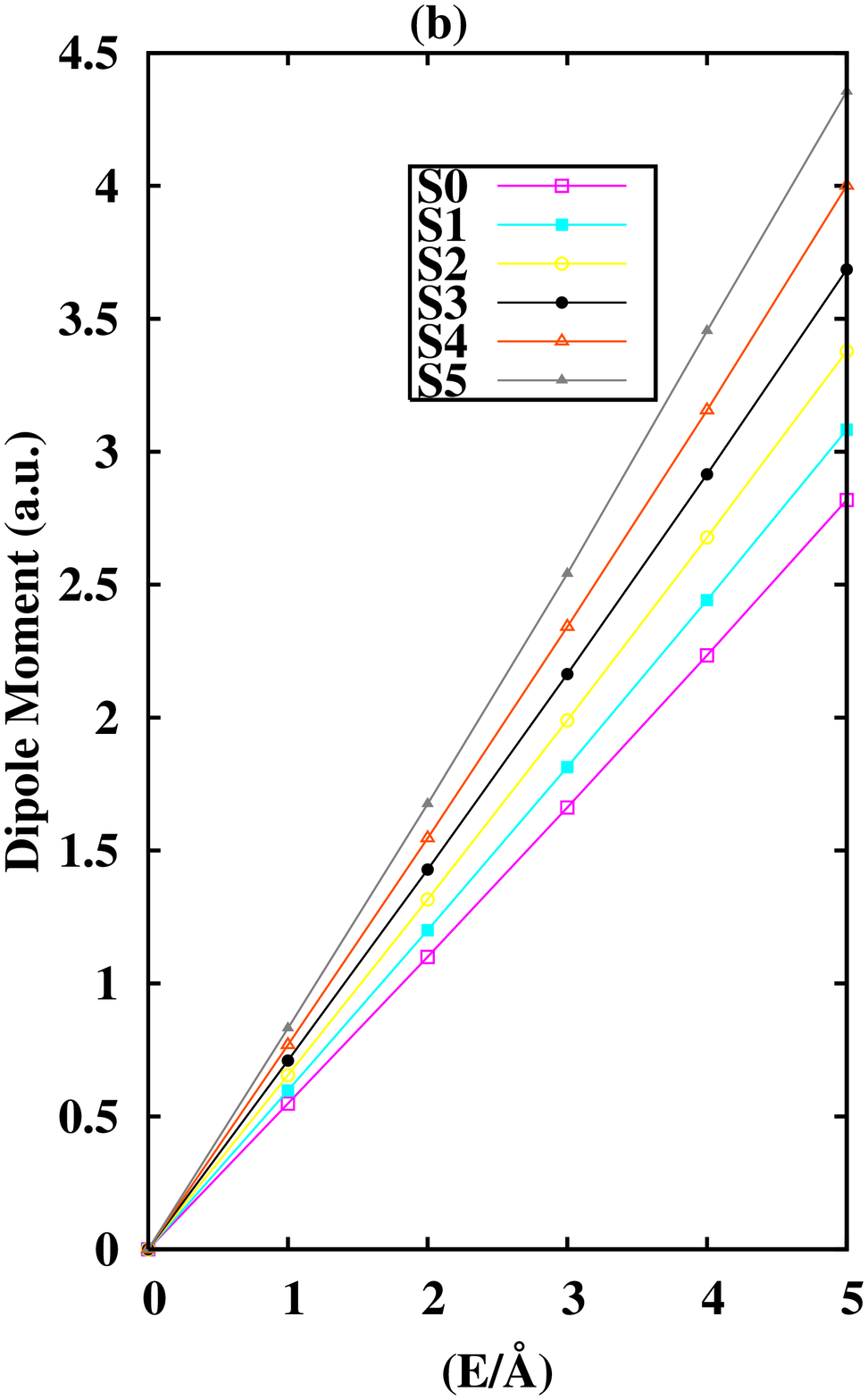}
\caption{(Color online) (a) The evaluation of the cohesive energy of different systems ($S_{d}\,d=0-5$) under external electric field. The horizontal solid line represents $E_{c}$ of bulk Si. Right panel (b) shows the dipole moment (in atomic units (a.u.)) of different systems under external electric field. }
\label{Energy-Filed}
\end{figure}

Figure~\ref{Energy-Filed}(a) shows the evaluation of the cohesive energy of different systems $S_{d}$ under external electric field $E$, which was applied in a direction perpendicular to the plane of silicene sheet, and distortions $S_{d}$. It is clear that $E_{c}$ increases with $E$. One can also judge that external distortion decreases $E_{c}$ of silicene. The pristine silicene has large $E_{c}$ at higher fields, which suggests that silicene may become unstable. However, if we distort silicene in such a way that changes the bond angle between sublattice A and B, then $E_{c}$ becomes smaller even smaller than bulk Si at higher strain. Though, distortions decrease $E_{c}$ of silicene, but it may not be dynamically stable due to imaginary frequencies in the phonons mode along the $\Gamma-K$ direction of BZ.\cite{Cahangirov}
The increment (decrement) of  $E_{c}$ in electric field (distortion) can be partially understood in terms of dipole-dipole interactions. Figure~\ref{Energy-Filed} (b) shows the dipole moments under electric field of different systems $S_{d}$. We see that electric field induces dipole moment in silicene which grows with the strength of electric field. Therefore, $E_{c}$ increases with electric field. The marked difference between $E_{c}$ and dipole moment is the behaviour of silicene in the external field-- the dipole moment increases linearly with electric field, consistent with the previous work,~\cite{RGA} but the cohesive energy does not follow a linear trend at higher fields. It is also noticeable that the distorted systems have larger induced dipole moment, and on the other hand, the distorted systems have lower $E_{c}$. So, the electronic polarization of valence electrons also contributes to the cohesive energy of silicene.

\begin{figure}[]
\includegraphics[width=0.15\textwidth]{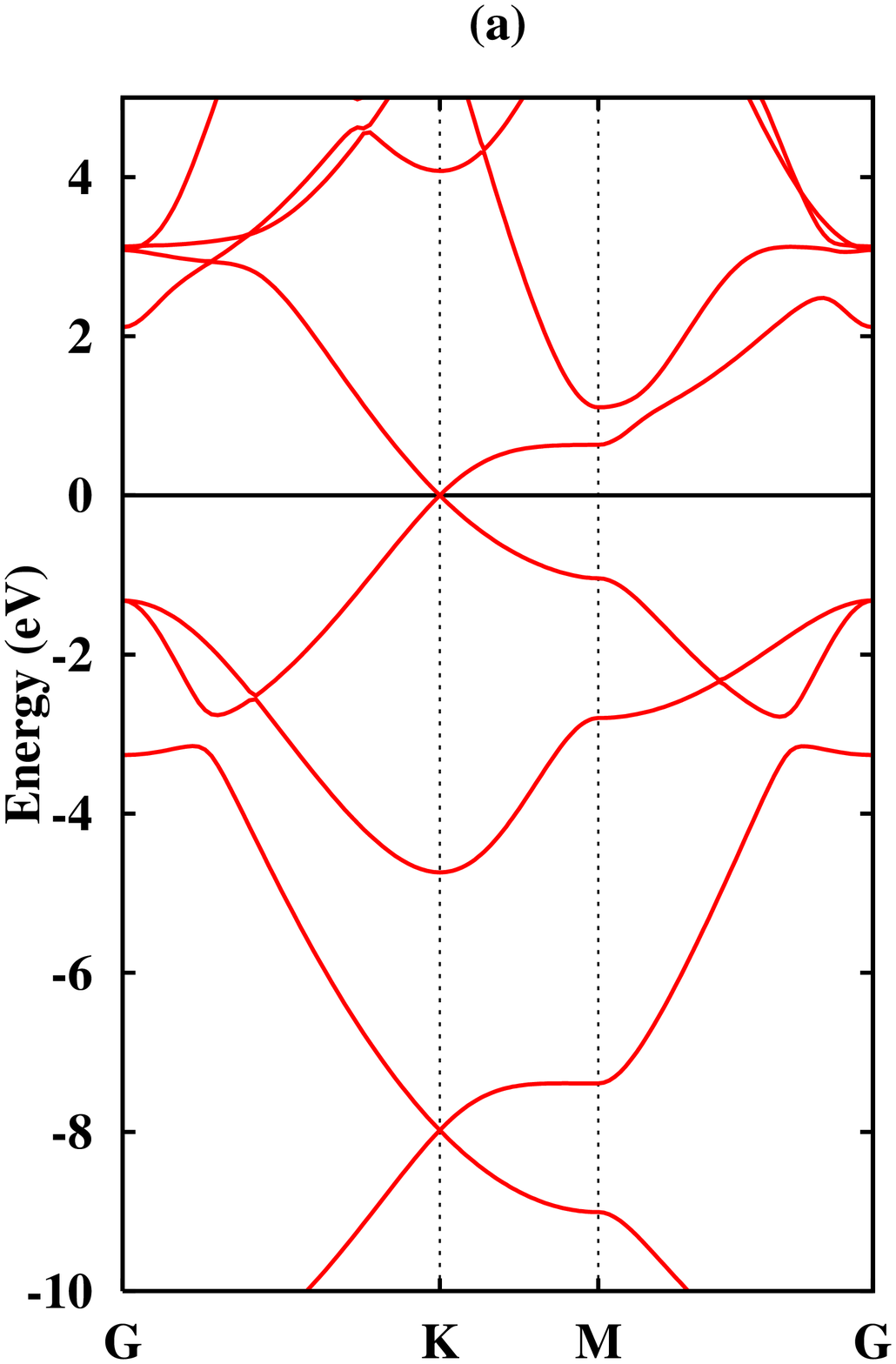}
\includegraphics[width=0.15\textwidth]{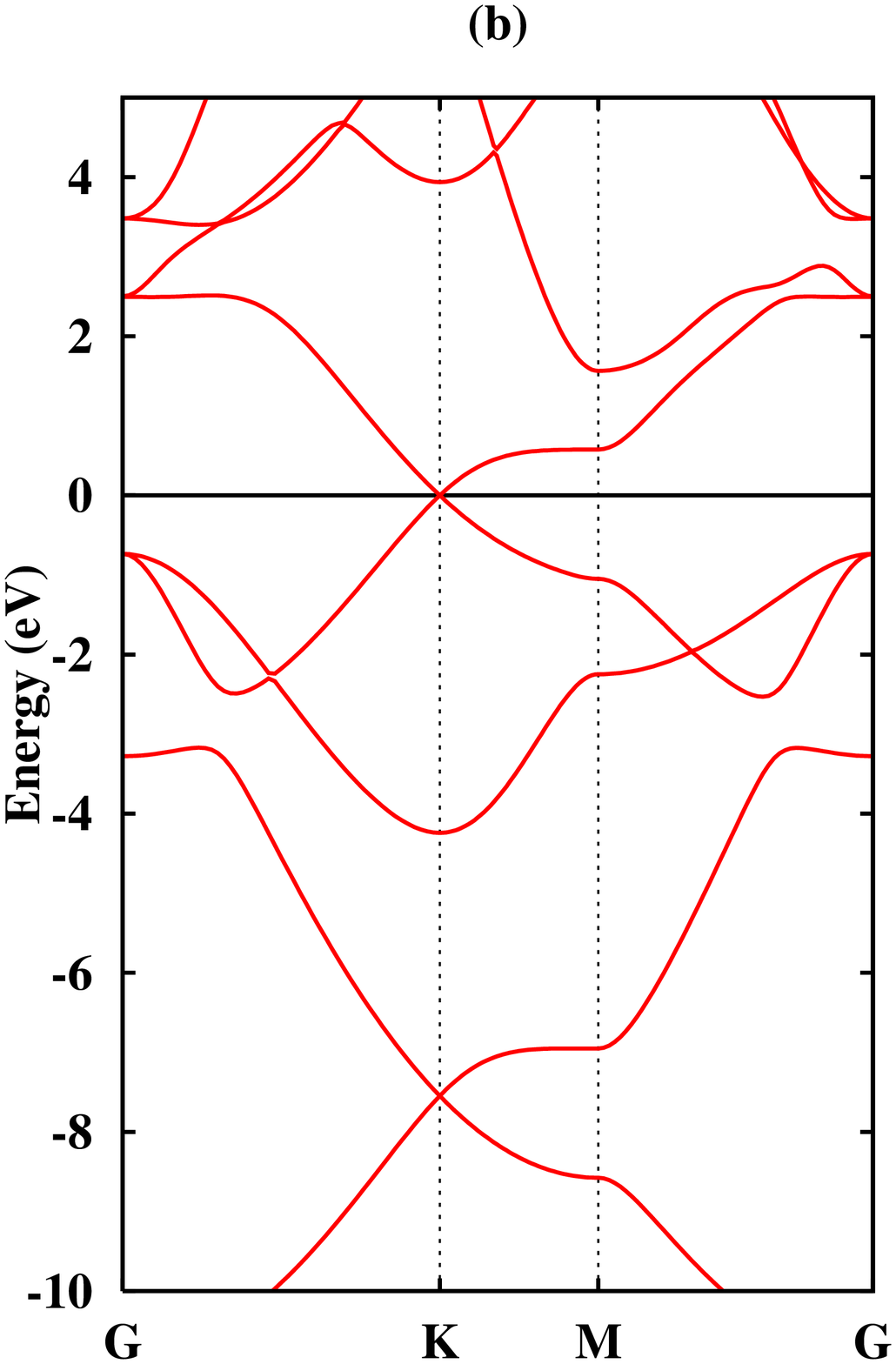}
\includegraphics[width=0.15\textwidth]{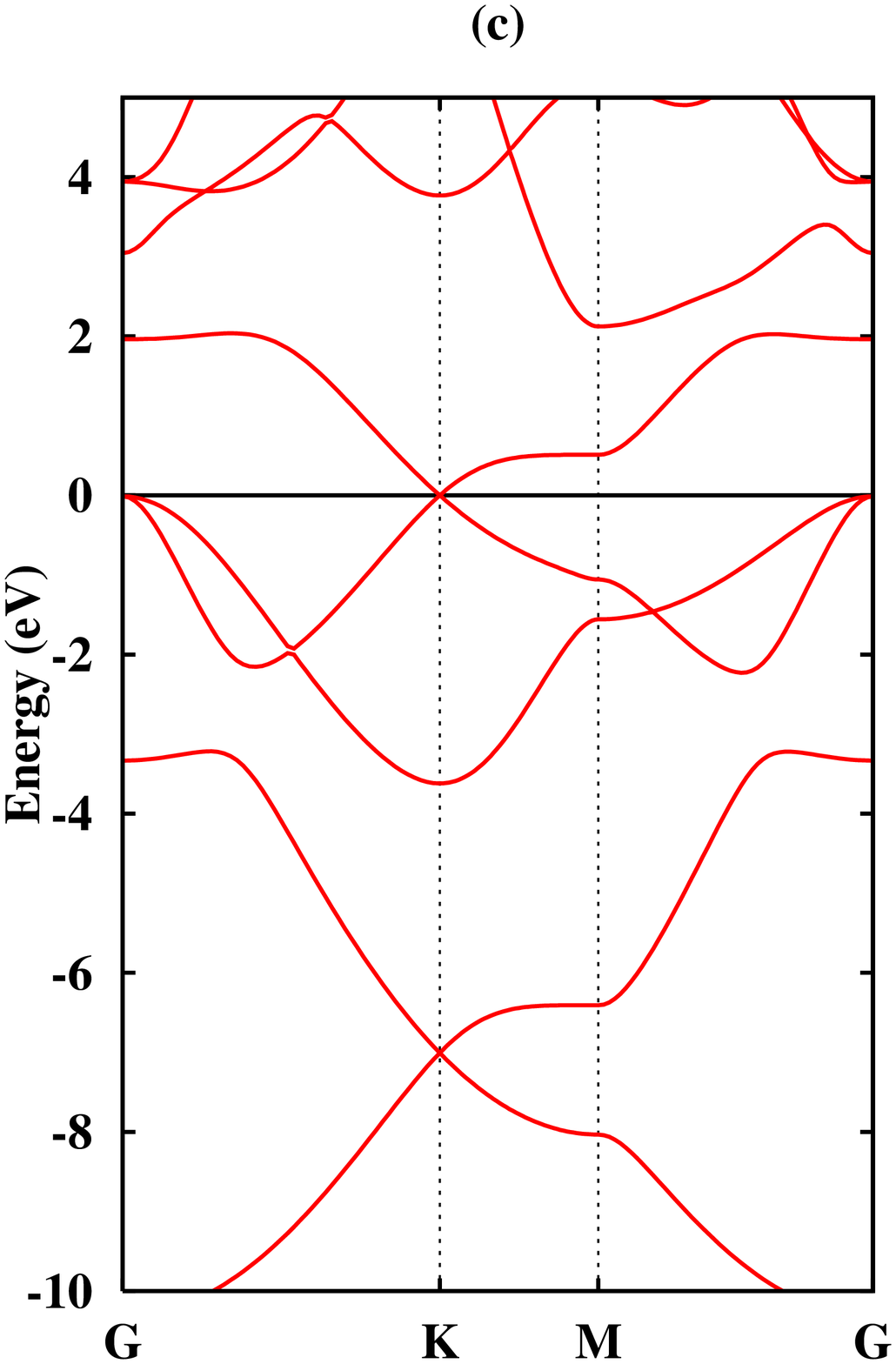}
\includegraphics[width=0.15\textwidth]{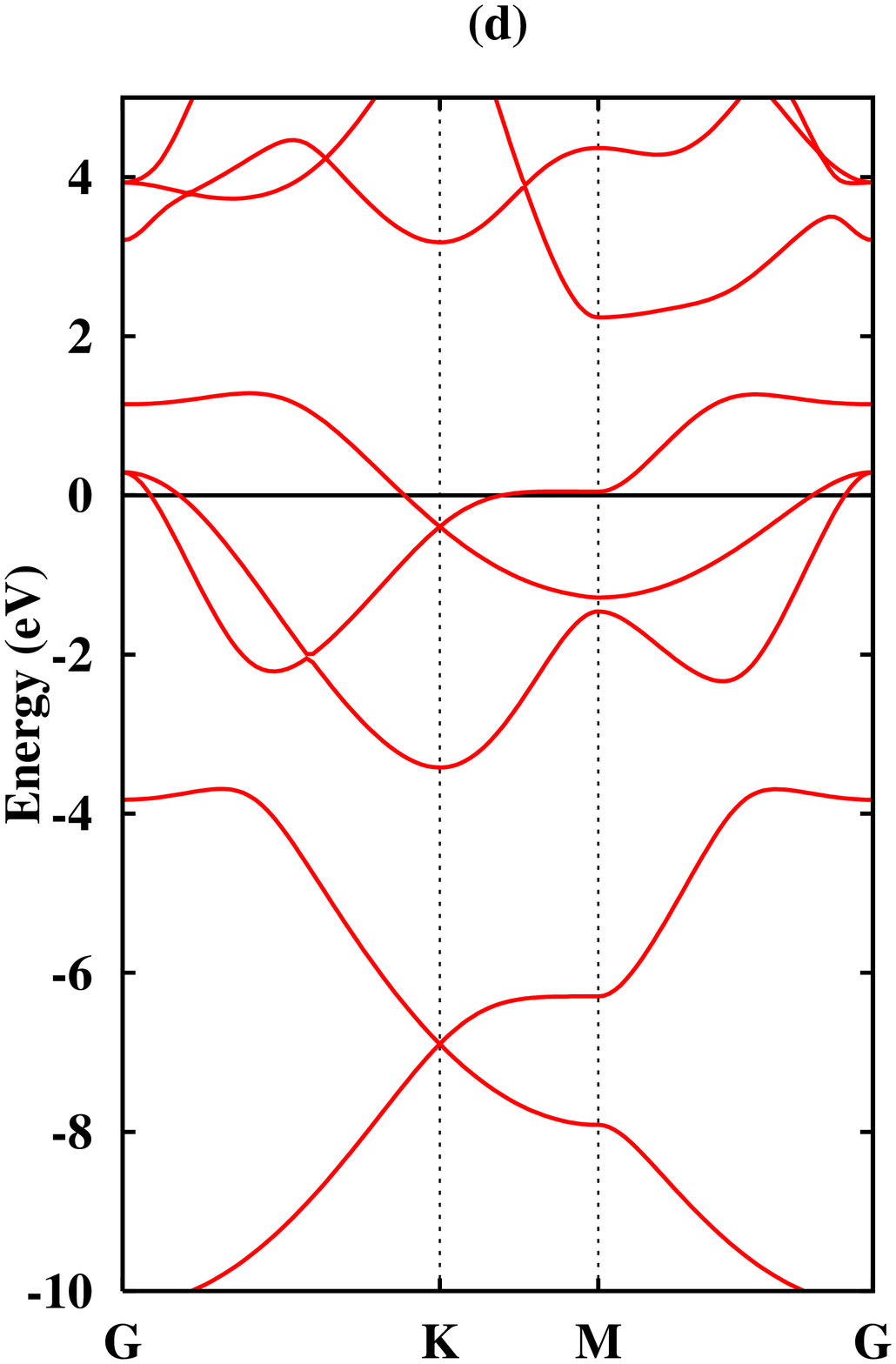}
\includegraphics[width=0.15\textwidth]{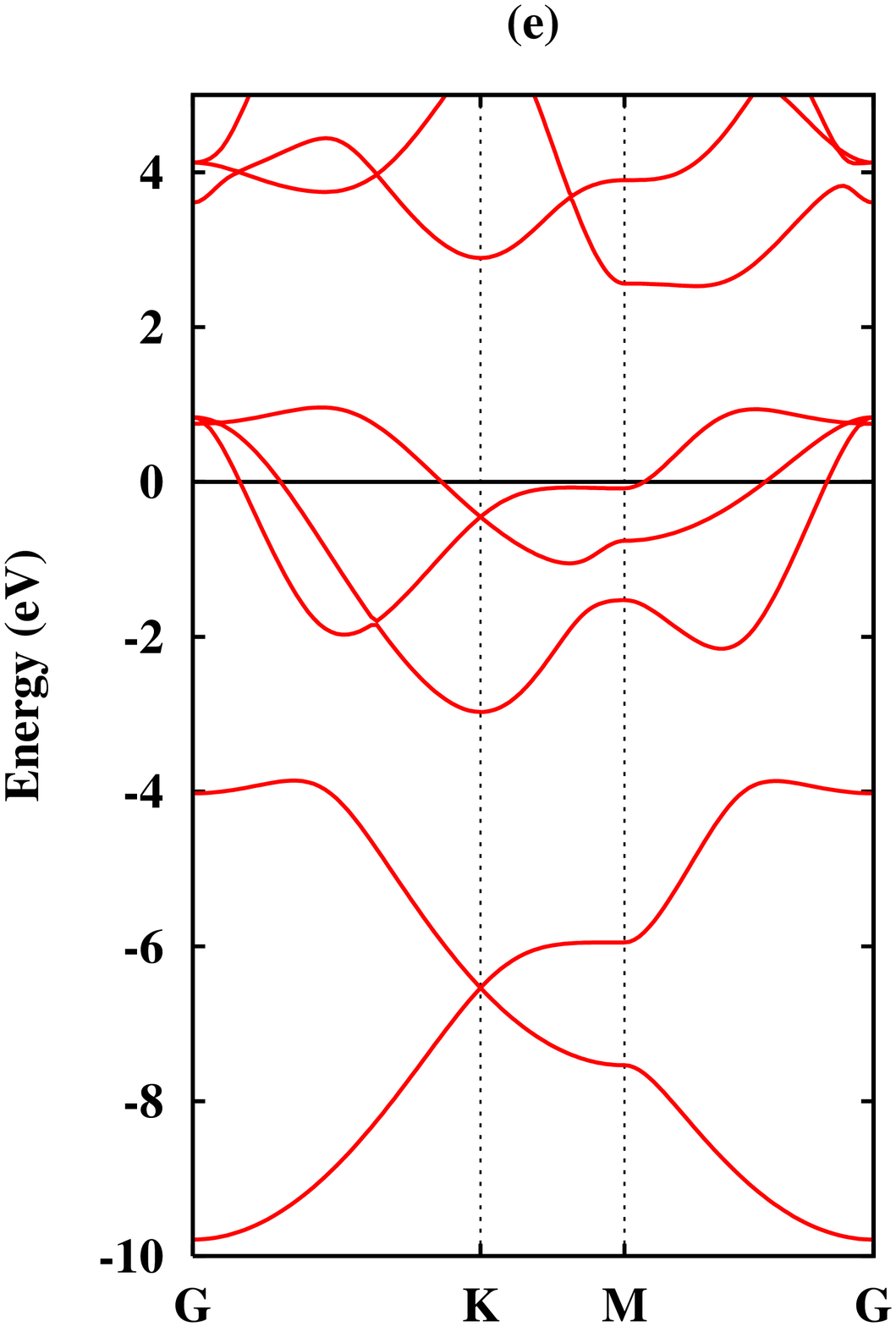}
\includegraphics[width=0.15\textwidth]{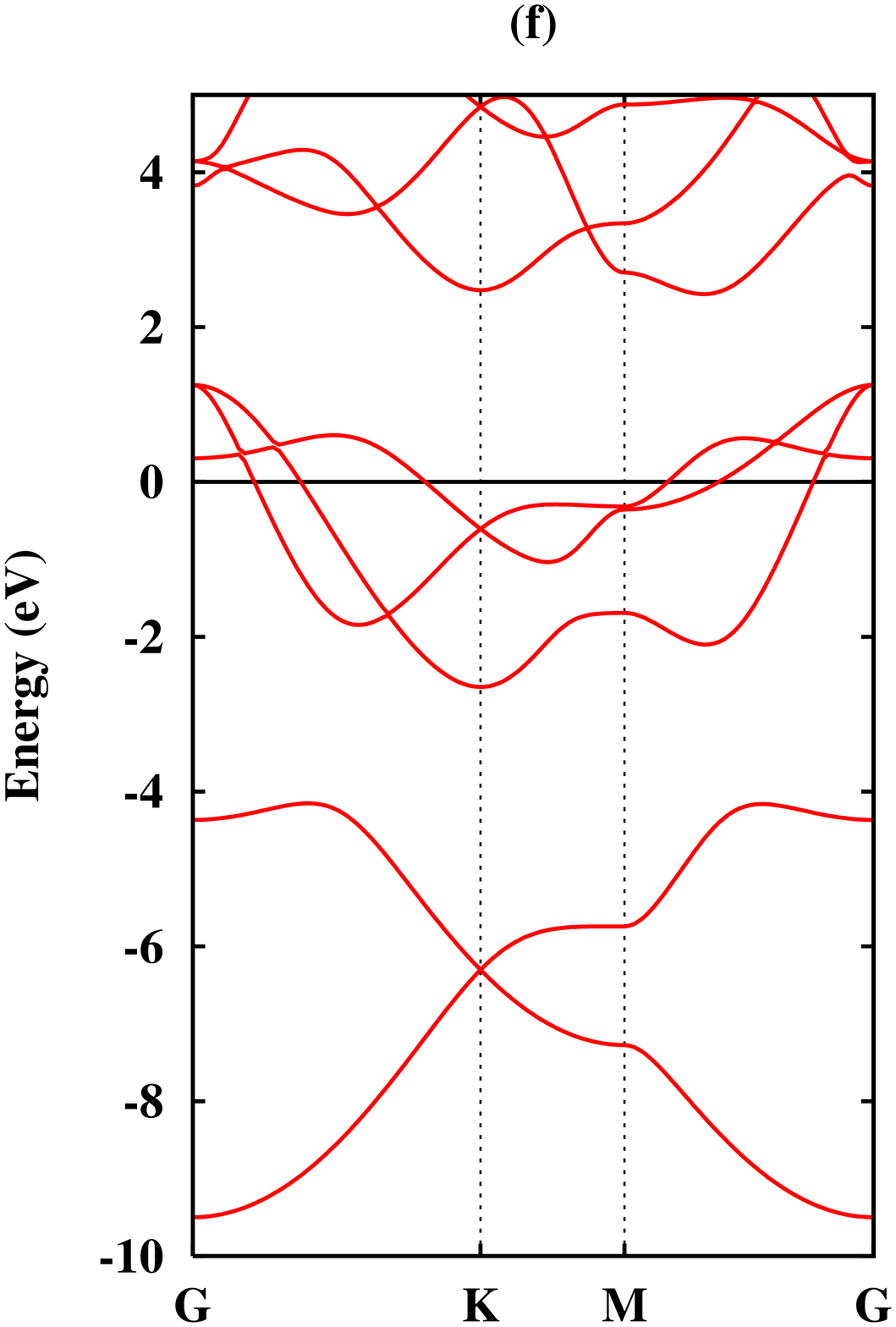}

\caption{(Color online) Electronic band structure of silicene in zero electric field of different systems $S_{d}$ . Labels (a)--(f) represent different systems $S_{0-5}$ correspond to different bond angles and lengths as mentioned in Table~\ref{angle}. The horizontal line shows the Fermi level which is set to zero.}
\label{Band-strain}
\end{figure}

To have some physical arguments for the behaviour of $E_{c}$ under electric field, and to know why $E_{c}$ changes with $E$ and distortion, we calculated the electronic band structures of silicene under different distortions $S_{d}$ and electric fields. Our calculated electronic structures are shown in Fig.~\ref{Band-strain}. The electronic structure of pristine silicene [Fig.~\ref{Band-strain}(a)] shows that it is a zero-band-gap semiconductor similar to graphene. The Dirac point is located at the $K$ point and coincides with the Fermi level $E_{F}$, consistent with the previous calculations.\cite{Cahangirov,Drum}

Figure~\ref{Band-strain} (b--f) also shows the band structure of silicene under different distortions. It is interesting to note that strain does not induce any band gap opening. This behaviour is consistent with the well studied system, graphene.\cite{choi,rakshit} When the structure of silicene is distorted, the linear dispersion relation at $K$ near the Fermi energy is preserved, and the Dirac point $E_{D}$ is shifted below the Fermi energy at higher distortions. Note that silicene retains the semi-metallic band structure, until the bond length (angle) $\le 2.37$ \AA (107.85$^{0}$), which is close to the bulk Si values. Major changes in the electronic structure around the $\Gamma$ point are also visible. The $\Gamma$ degenerate band around $-2$ eV also retains its degenerate behaviour under the studied distortions. However, when the silicene is distorted this degenerate band moves towards the Fermi energy and finally touches the Fermi energy before transferring to metallic silicene. Similarly the high energy bands around $2$ eV also move towards the Fermi energy at higher strains. So, as we increase the bond angle, the lowest conduction band of silicene near $\Gamma$-point drops and is filled. At higher strains, all bands cross the Fermi energy and silicene is no longer semi-metallic. It suggests  that silicene can absorb such small deformations before changing its electronic structure. Such behaviour of silicene advices that silicene can be used in nano devices provided that one should not cross this distortable limit. From these calculations, we can infer that the bond angle(length) between the Si atoms in silicene plays a vital role to control the Dirac cones. We believe that the experimentally determined scattered results~\cite{ZrB2,Ag1, Chen, Lin} may not only be due to the interface effect,\cite{peng} but also due to different distortions, which change the bond angle of silicene, when deposited on a substrate. 

Even distortion changes the electronic structure, and at higher strain (bond angles smaller  than $\sim 107^{0}$) the Dirac point moves below the Fermi energy and retains its linear dispersion. The location of the Dirac point is very important and it can be used to engineer the carrier (holes or electrons) concentration $\rho$. For a 2D Dirac system, $\rho$ can be estimated from the distance between $E_{D}$ and $E_{\rm{F}}$, using the expression $\rho=sgn(E_{\rm{D}}-E_{\rm{F}})((E_{\rm{D}}-E_{\rm{F}})^{2})/\pi\beta^{2}$,
~\cite{Lazzeri} where $\beta$ can be calculated from the linear dispersion around $K$ as $E_{K+k}=\pm\beta\,k$. Interpretation of Fig.~\ref{Band-strain} in terms of $\rho$ shows that silicene  behaves as a $n$-type doped system beyond system $S_{2}$, $E_{D}$ is shifted below the Fermi energy and $\rho$ is negative for the distorted silicene. This suggests that such distortion can behave as a self doping in silicene. The physical mechanism of this behaviour can be understood from the band structure (see Fig.~\ref{Band-strain}), where the highest occupied states at $\Gamma$ point move towards the Fermi energy under different strains. At bond angle $\sim 102^{0}$ ($S_{3}$)(compressed about 11\%), the occupied states (strain free case) become unoccupied and cross the Fermi energy and silicene behaves as a $n-$type doped system. We see from these band structures that the band structure of silicene under small strain follows the linear dispersion relation at $K$ point. Such strains can only lower the Dirac point, \text{i.e.}, changing the carriers type.  Strain is a good tool to engineer the band structure of silicene. Chemical doping, e.g., Li, C, H, Co, dramatically changes the band structure and silicene does not have  a linear dispersion relation around $K$ point in the BZ. Such chemical doping may either transform silicene to metallic or insulating one.  
It seems that the electronic structure can be tuned either by strain or electric field.
Indeed, large strain or E-field can destroy the linear relation of dispersion.

\begin{figure}[]
\includegraphics[width=0.15\textwidth]{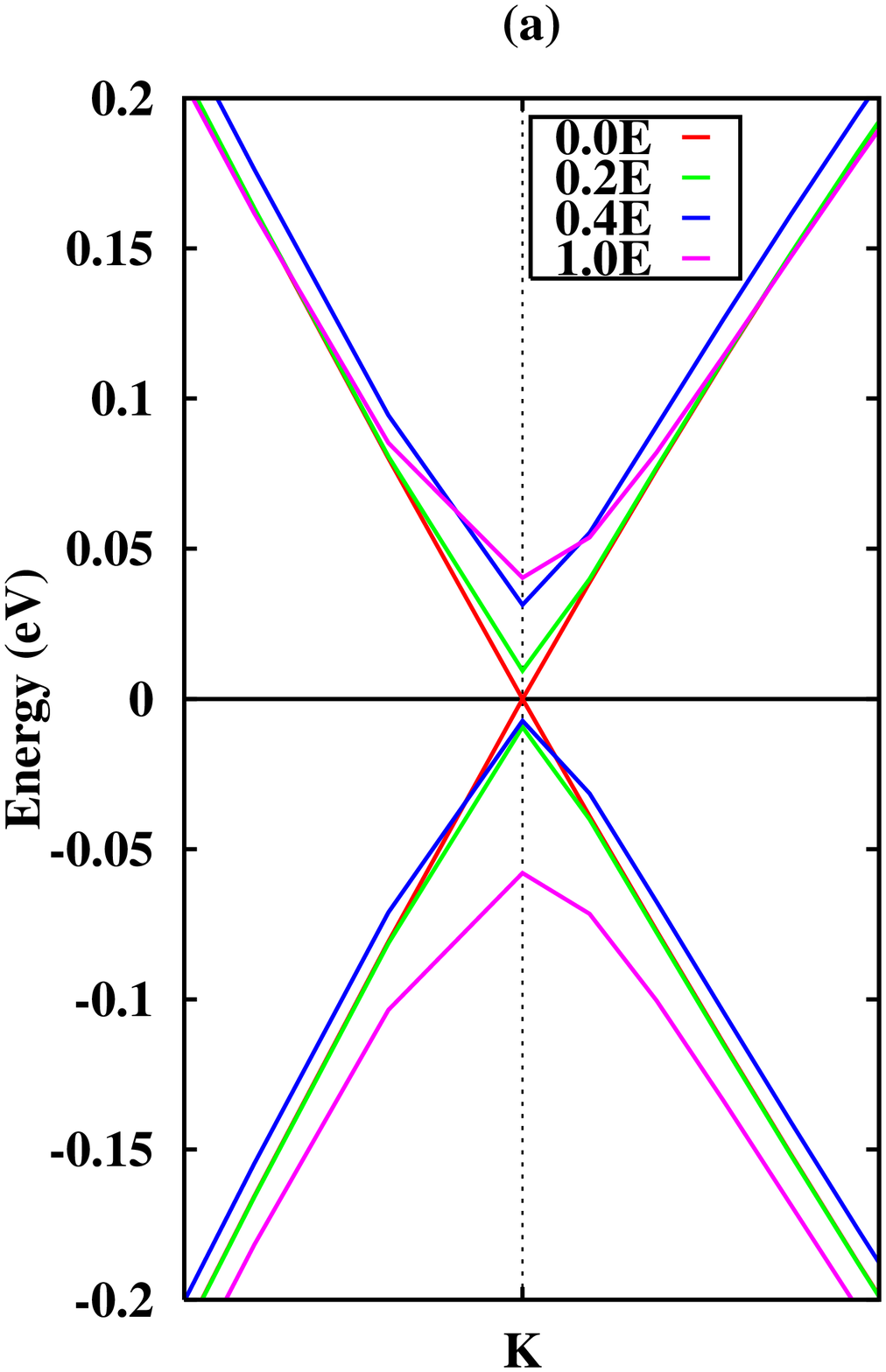}
\includegraphics[width=0.15\textwidth]{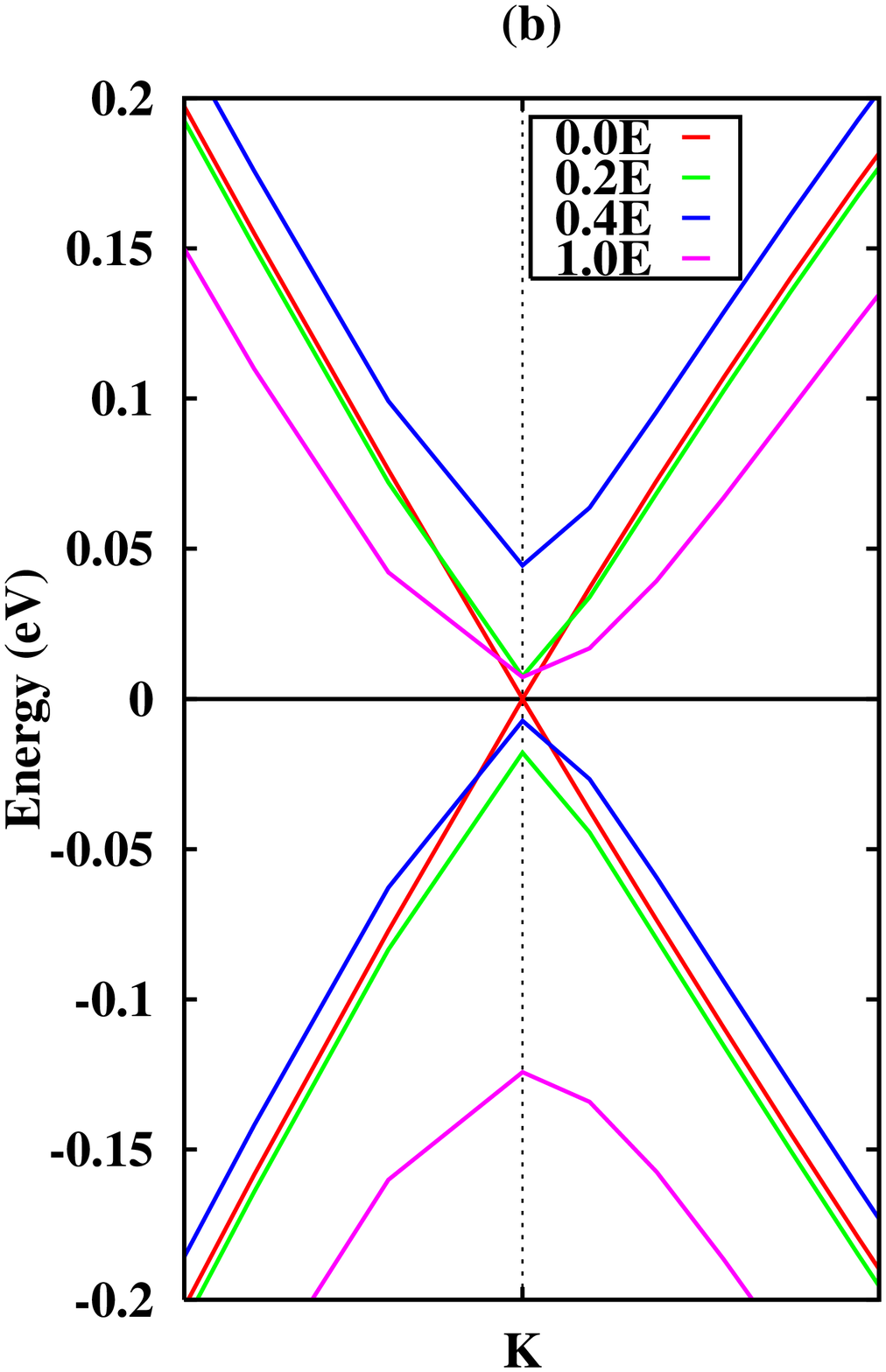}
\includegraphics[width=0.15\textwidth]{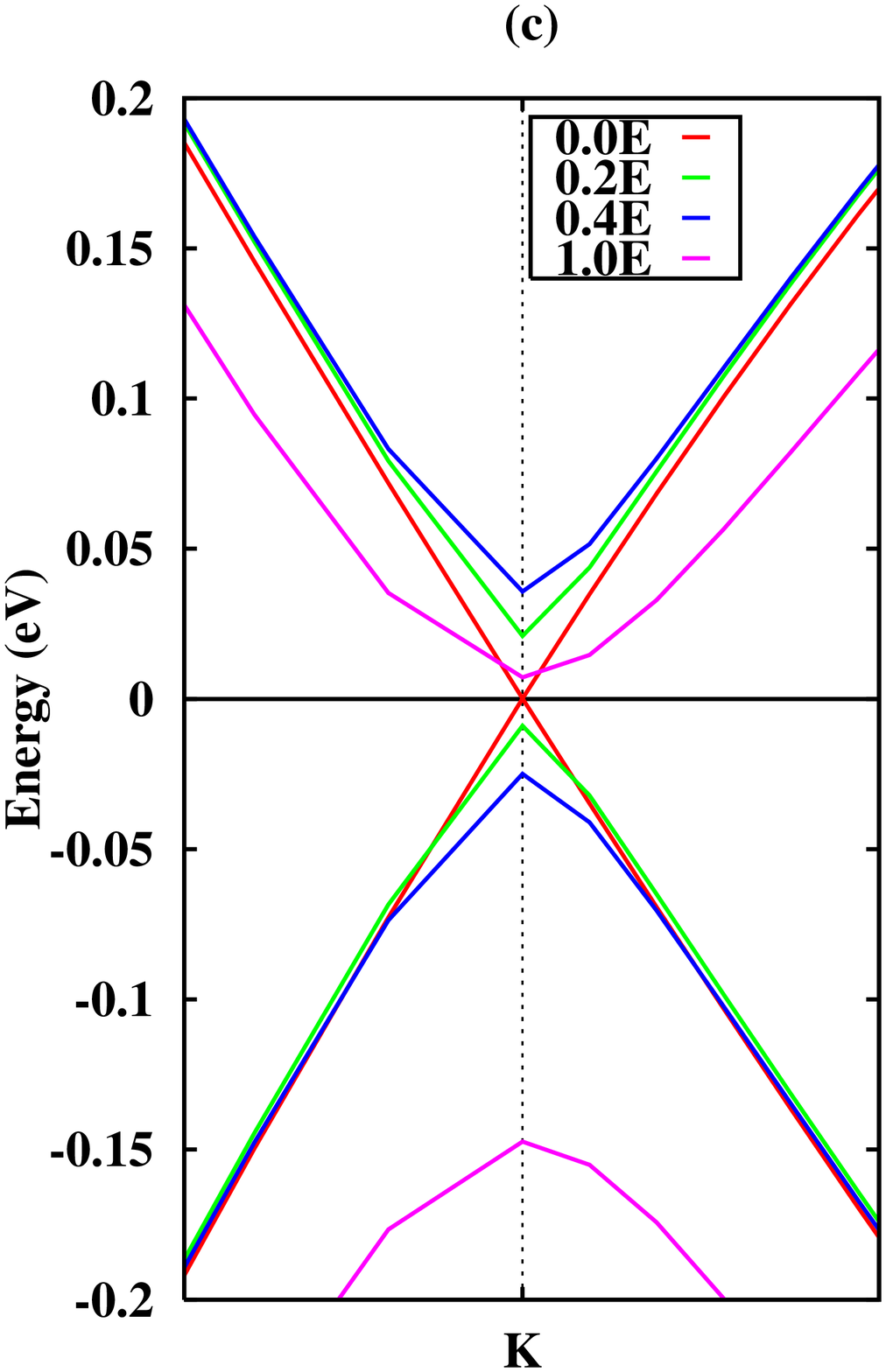}
\includegraphics[width=0.15\textwidth]{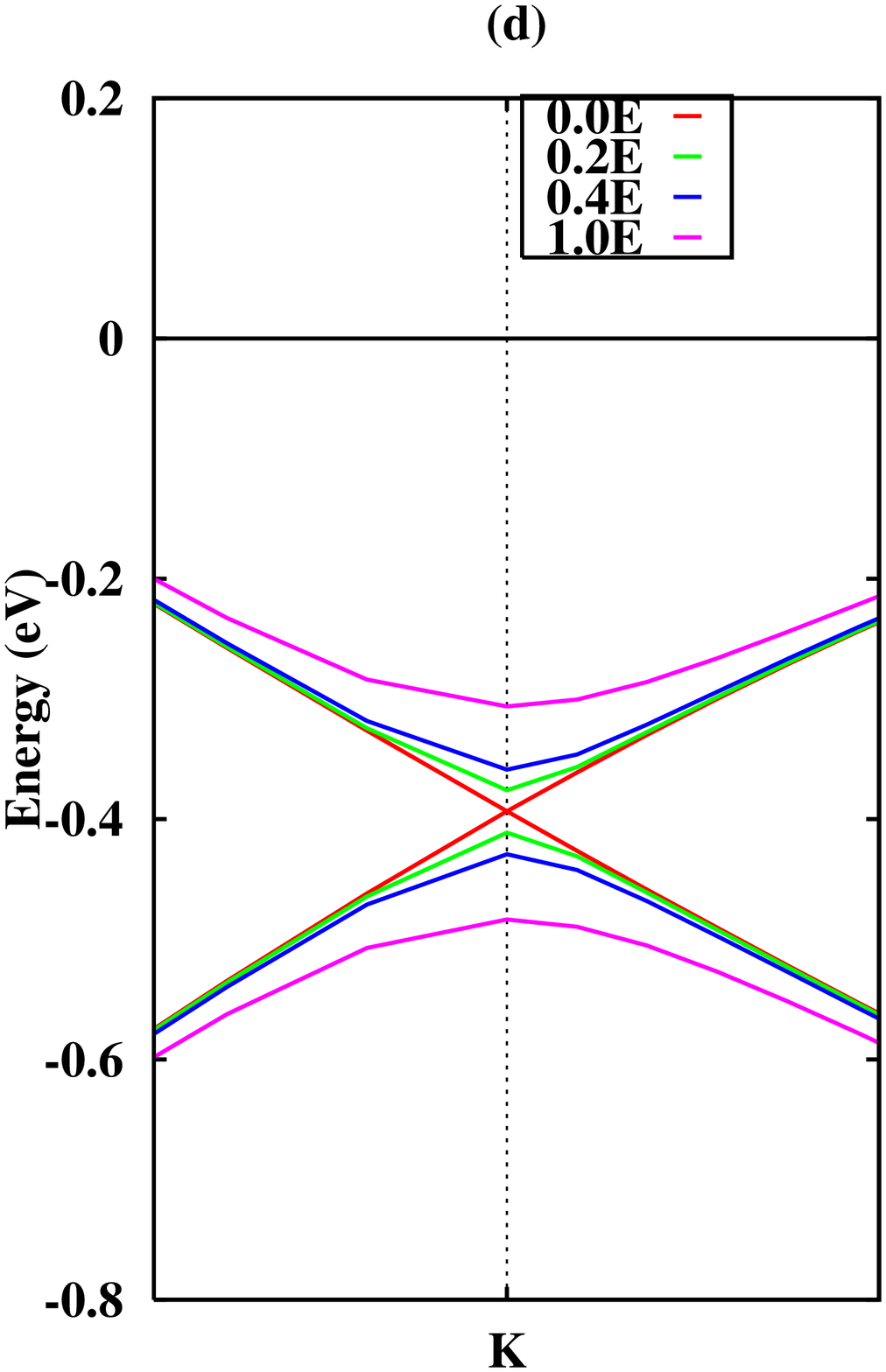}
\includegraphics[width=0.15\textwidth]{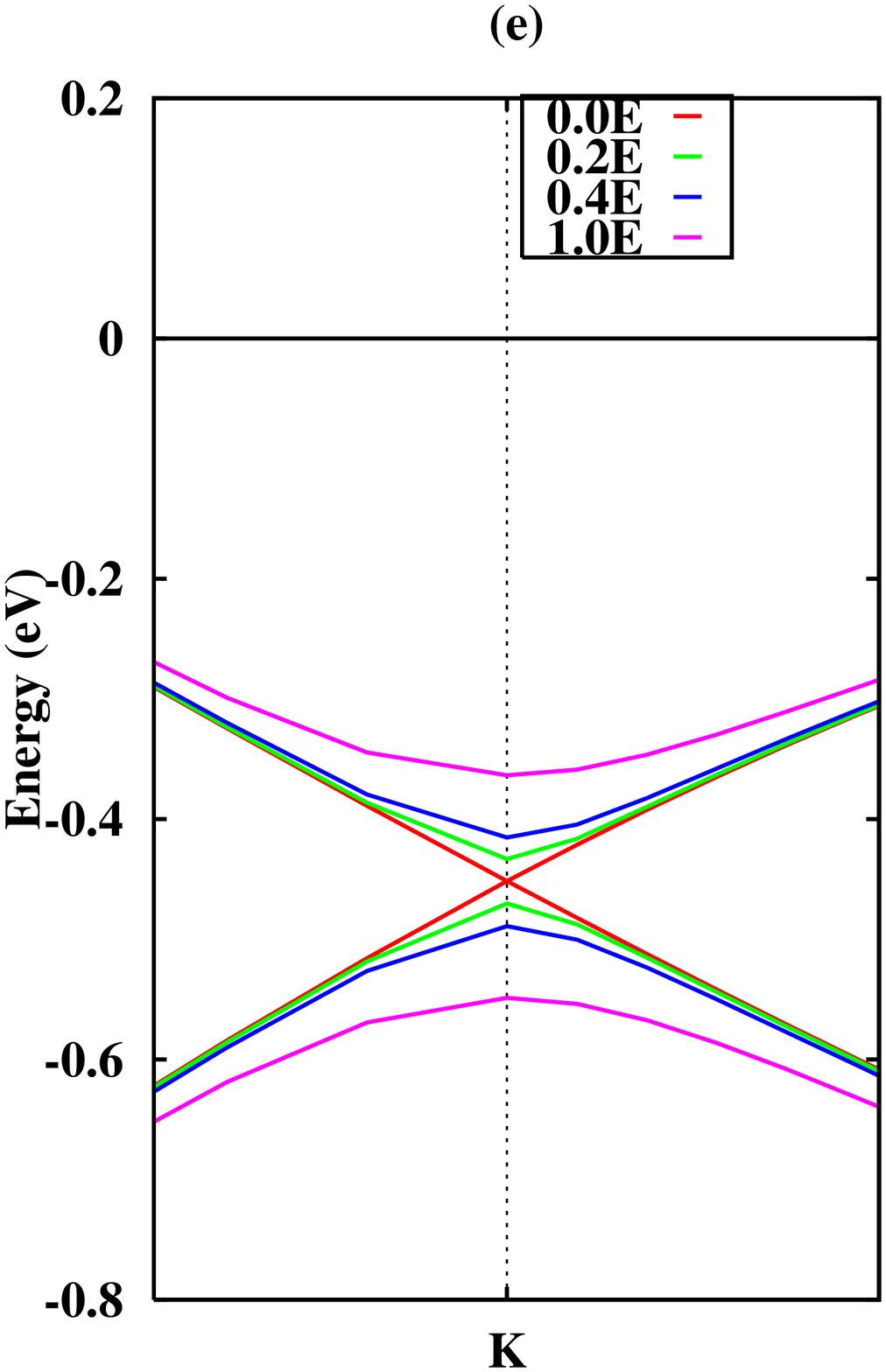}
\includegraphics[width=0.15\textwidth]{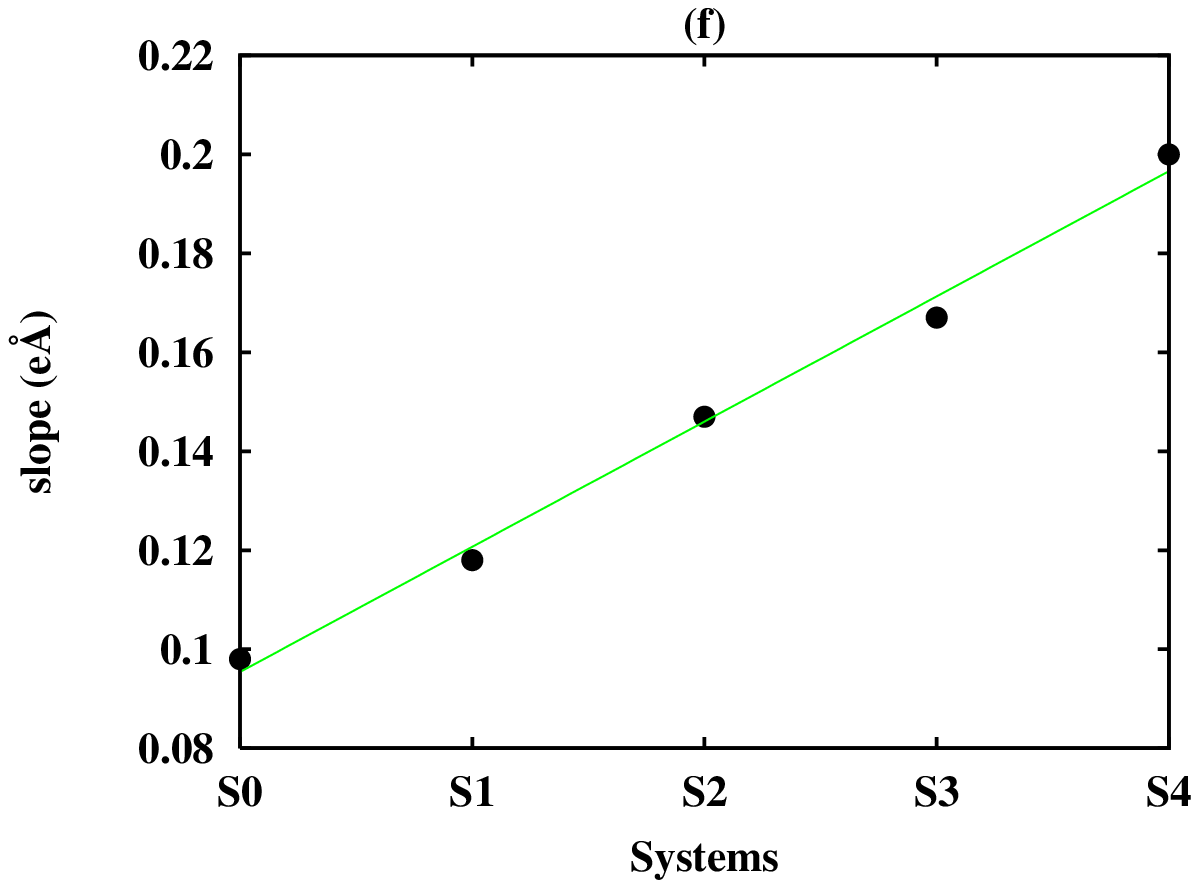}

\caption{(Color online) Band structure (Dirac points) of silicene under different strain and E-field.  Labels (a)--(e) represent different systems $S_{0-4}$ correspond to different bond angles and lengths as mentioned in Table.~\ref{angle}  The horizontal line shows the Fermi level which is set to zero. Whereas figure labelled (f) shows slope $\gamma$ of different systems $S_{d}$ }
\label{Band-Efield}
\end{figure}

Now we investigate the electronic structure of silicene in the presence of external electric field $E$. The calculated band structure of silicene around Dirac point for various electric field and distortions $S_{d}$ are shown in Fig.\ref{Band-Efield}.
The band gap in silicene is opened due to breaking of inversion symmetry by the electric field since the potential seen by the atoms at the sites $i$ and $j$ are different. In this situation, the
finite value of on-site energy difference $\Delta$ arises due to the potential difference and hence, we
can write $\Delta=\alpha(V_{i}-V_{j})$, where $\alpha$ and $V_{i}(V_{j}$)  are proportionality constant and the
potential seen by the atom at the site $i(j)$, respectively. For constant electric field between the sheets, the potential difference $\Delta V=V_{i}-V_{j}=eE\delta=F\delta$, where $F$ is the electric field intensity and $\delta$ is the buckling parameter of silicene, which is non-zero (zero) for silicene (graphene). This finite value of $\delta$ in silicene breaks the symmetry of silicene in the presence of electric field and helps in opening up a band gap in silicene.

We also found that the band gap $E_{g}$ increases linearly with electric field $E$ (not shown here), similar to Fig.~\ref{Energy-Filed} (b). It will be very essential to estimate this proportionality constant $\gamma$ ($E_{g}=\gamma\,F$), which can be estimated from the slope of $E_{g}$ vs $E$. For this purpose, we varied $E$ from 0.0 to 1.0 $V/$\AA\, and  calculated $E_{g}$ at $K$ point for each value of $E$. Our estimated $\gamma$ for pristine silicene was $0.098\,e$ \AA,\,  which is comparable with the previous values.~\cite{ZNi,Drum}
Figure \ref{Band-Efield}(f)  shows $\gamma$ of all systems. It shows that $\gamma$ also increases linearly with distortion irrespective of the location of Dirac point. This is interesting to note that the curvature of bands decreases almost inversely with $E$, and at higher $E$, the dispersion around $K$ becomes parabolic, consistent with the tight binding approximation.\cite{ZNi} It indicates the global linearity of effective masses with electric field. Hence, the band gap and
the effective masses are proven to be proportional to $E$.

{Before summarizing our work, we must note that the band gaps and the band widths are usually underestimated by DFT-LDA, which can be corrected by GW type calculations.\cite{shu,wei,asia} Previous GW calculations show $\sim$ 50 to 140\% enhancement of band gap as compared with DFT-LDA calculations.\cite{wei,asia} Therefore, it is speculated that our calculated E-field induced band gaps and band widths may be larger than those shown in Fig. \ref{Band-strain} and \ref{Band-Efield}. In the light of previous GW calculations,\cite{shu,wei} the $\pm 0.40$ eV width of the linear regime\cite{Ag1,Chen}, the corrected DFT picture may not be as far from the measurements. Nevertheless, both GW and LDA predict linear dispersion relation around $K$-point and semi-metallic behaviour in low buckled  silicene.\cite{shu} }

In summary, we used DFT to elucidate the origins of experimentally determined different Dirac points in silicene. We showed that the cohesive energy of silicene decreases with distortions and increases with external electric fields. The band structure calculated at different distortions and electric fields showed that as silicene was distorted, the cohesive energy decreased due to unoccupied states at $\Gamma$, and at larger distortions, the Dirac cones also shifted below $E_{F}$ and silicene behaved as an $n$-doped silicene. This $n$ type behaviour of silicene also participated in lowering of the cohesive energy of silicene. From our DFT-LDA calculations, we came to the conclusion that silicene can retain the semi-metallic band structure when the bond angle between the Si atoms in silicene is closer to the bond angle  between the Si atoms in bulk Si. Therefore, the scattered experimental reports about the position of Dirac point may be attributed to different distortions of silicene on different substrates. We believe that it is not only the interfacial effect, but also the bond angle between the Si atoms in silicene that destroys the Dirac cones in silicene. These two factors can determine the existence or nonexistence of semi-metallic nature of silicene, when deposited on  a substrate.

\section*{ACKNOWLEDGMENTS}
G.R. acknowledges the cluster facilities of NCP, Pakistan. The author is grateful for insightful discussions with {V\'{\i}ctor M. Garc\'{\i}a-Su\'arez} and Juliana M. Morbec.

\end{document}